\documentclass{nature}

\usepackage{graphicx}

\title{Magnetic order in BaFe$_2$As$_2$, the parent compound of the FeAs based superconductors in a new structural family}

\author{Q. Huang$^{1}$, Y. Qiu$^{1,2}$, Wei Bao$^{3}$, M.A. Green$^{1,2}$, J.W. Lynn$^{1}$, Y.C. Gasparovic$^{1,2}$, T. Wu$^{4}$, G. Wu$^{4}$ \& X. H. Chen$^{4}$}

\begin{document}

\maketitle

\begin{affiliations}
\item NIST Center for Neutron Research, National Institute of Standards
and Technology, Gaithersburg, MD 20899, USA
\item Department of Materials Science and Engineering,
University of Maryland, College Park, MD 20742, USA
\item Los Alamos National Laboratory, Los Alamos, NM 87545, USA
\item Hefei National Laboratory for Physical Science at Microscale and Department of Physics, University of Science and Technology of China, Hefei, Anhui 230026, China
\end{affiliations}

\begin{abstract}
In addition to higher superconducting transition temperatures (\mbox{\boldmath$T_C\approx 55$} K)  compared with the single layered cuprates\cite{A042053,A042105,A042582,A044290}, the newly discovered
iron oxypnictide superconductors offer additional compositional variation.
The family of FeAs based superconductors with the same ZrCuSiAs-type (1111-type) structure now include \mbox{\boldmath$Ln$}FeAsO, \mbox{\boldmath$Ln$}=La, Sm, Ce, Nd, Pr, Gd, Tb or Dy\cite{Kamihara2008,A033603,A033790,A034234,A034283,A034384,A042053,A042105,A042582,A043727,A044290,A060926}.
In a similar fashion to the CuO$_2$ layers present in cuprates,
the FeAs layers have been shown in theory to dominate the electronic states that produce superconductivity\cite{A030429,A031279,A031282}. Cuprate superconductors distinguish themselves structurally by adopting different stacking sequences of the CuO$_2$ and electronically inactive ``spacer'' layers. Using the same structural philosophy, layered materials  with the formula (\mbox{\boldmath$A$},K)Fe$_2$As$_2$, \mbox{\boldmath$A$}=Ba or Sr, which crystallize in a different ThCr$_2$Si$_2$-type (122-type) structure, have recently been reported and possess a \mbox{\boldmath$T_C$} as high as 38 K\cite{A054630,A061209,A061301,A061459}. Here, we report the neutron diffraction studies of BaFe$_2$As$_2$ that show, in contrast to previous studies on the 1111-type materials\cite{A040795,A043569,A061450,A062195}, a phase transition to a long-ranged antiferromagnetic state at the same temperature where a structural transition from tetragonal to orthorhombic symmetry also occurs. The discontinuity in the intensity of the (220)\mbox{\boldmath$_T$} Bragg reflection implies a first-order transition as a result of the greater flexibility within the nonequivalent layers in the 122-type structure. This is in contrast to the separate second-order structural and magnetic transitions seen for the previously reported 1111-type materials\cite{A040795,A043569,A061450,A062195}. Although the magnetic and structural transitions occur differently in the BaFe$_2$As$_2$ and the 1111-type materials, this work clearly demonstrates that the complete evolution to a low symmetry structure is a pre-requirement for the common iron magnetic order to occur.

\end{abstract}

Like the parent compounds $Ln$FeAsO of the 1111-type FeAs superconductors, the 122-type parent compounds BaFe$_2$As$_2$ and SrFe$_2$As$_2$ are not superconducting and show a similar pronounced anomaly in resistivity at $T_S\sim 140$ K\cite{A054021,A054630,A061459} and 205 K\cite{A061209,A061301,A061043}, respectively.
Structural refinements have been performed only for BaFe$_2$As$_2$ using X-ray powder diffraction, and a second-order structural phase-transition at $T_S$ has been concluded like in LaFeAsO\cite{A054021}. The $^{57}$Fe M\"{o}ssbauer spectra have been measured at 298, 77 and 4.2 K
for BaFe$_2$As$_2$, and long-range magnetic order exists at 77 and 4.2 K\cite{A054021}. No such
spectroscopic studies have yet been performed for SrFe$_2$As$_2$. Careful specific heat measurements, however, indicate the phase transition at $T_S\approx 205$ K in SrFe$_2$As$_2$ as a first-order one\cite{A061043}. So far, no magnetic structure in the new 122-type materials has  been reported, neither is it clear whether the magnetic transition concurs with the structural transition at $T_S$ or is a separate phase transition as in LaFeAsO\cite{A040795,A043569} or NdFeAsO\cite{A061450,A062195}.

We synthesized 2 g of polycrystalline BaFe$_2$As$_2$ sample using the solid state reaction as described in Ref.~[19]. The resistivity was measured using the standard four-probe method while the sample was cooled. As shown in Fig.~1, the drop of resistivity at $T_S\approx 142$ K is sudden and steep, as reported by Krellner et al.\ at 205 K for SrFe$_2$As$_2$\cite{A061043}.
Neutron powder diffraction spectra were measured with neutrons of wavelength $\lambda=2.079\AA$, using the high resolution powder diffractometer BT1 at the NIST Center for Neutron Research (NCNR).
The sample temperature was controlled by a closed cycle refrigerator. The spectrum measured at 175 K is shown in Fig.~2(a) together with the refined profile obtained with the GSAS program\cite{gsas}. Consistent with previous studies\cite{A054021}, the high temperature structure is well accounted for
by the tetragonal ThCr$_2$Si$_2$-type structure, and the
structural parameters from the refinement using space group $I4/mmm$ are listed in Table 1(a).

The neutron powder diffraction spectrum measured at 5 K is shown in Fig.~2(b). Our neutron data confirm the orthorhombic structural distortion previously reported in the X-ray study\cite{A054021}, which splits the (220)$_T$ Bragg peak of the $I4/mmm$ tetragonal unit cell, shown in the inset to Fig.~2(a), into two Bragg peaks (400)$_O$ and (040)$_O$ of the $Fmmm$ orthorhombic unit cell as marked in Fig.~2(b).
To clarify the nature of the structural transition at $T_S\approx 142$ K, we followed the diffraction intensity at $2\theta=95.8^o$ as a function of temperature. Above $T_S$, this angle corresponds to the peak position of the (220)$_T$; below $T_S$, it corresponds to the valley between the (400)$_O$ and (040)$_O$ peaks, see inset to Fig.~1. The intensity changes in an abrupt first-order fashion as shown in Fig.~1. Additionally, hysteresis was observed during a cooling and warming cycle with a temperature change rate of 15 K/h. Therefore, while the pronounced anomaly in resistivity is associated with the tetragonal-to-orthorhombic structural transition in both the 122 and 1111-types of the FeAs based materials, the structural transition in the newer 122-type material to a low symmetry structure occurs in a first-order transition.

Additional magnetic Bragg peaks are apparent in the neutron diffraction patterns measured below $T_S$. In the inset to Fig.~2(b), magnetic Bragg peaks from the 5K spectrum are highlighted by the vertical lines and are indexed using the orthorhombic unit cell. Irreducible representational analysis for BaFe$_2$As$_2$ in the $Fmmm$ space group shows magnetic moments are allowed along all three axis. Furthermore, the first order nature of the transition allows for any combination of these moments to contribute to the overall magnetic structure. Analysis of the magnetic Bragg reflections reveals a simple antiferromagnetic structure within the eight Fe ions in the orthorhombic unit cell. The magnetic and crystal parameters determined in a combined structural and magnetic refinement of the 5 K spectrum in Fig.~2(b) are listed in Table 1(b), and the magnetic structure
of BaFe$_2$As$_2$ is depicted in Fig.~3. The magnetic structure of BaFe$_2$As$_2$ is identical to that of LaFeAsO\cite{A040795}. This demonstrates a commonality of the magnetic interactions within the Fe layers and variation in the stacking of the layers, and therefore the interactions along the c axis, do not contribute greatly to the overall magnetic property.

The ordered magnetic moment 0.87(3) $\mu_B$ per Fe at 5 K in BaFe$_2$As$_2$ is substantially larger than the saturated moment 0.36(5) $\mu_B$ per Fe in LaFeAsO\cite{A040795}, but is comparable to the iron moment 0.9(1) $\mu_B$ in a different combined Fe and Nd antiferromagnetic structure in NdFeAsO\cite{A062195}. It should be noted that the Fe antiferromagnetic order occurs in the orthorhombic crystal structure below $T_S$, and the $a$ and $b$ axes are not equivalent. For LaFeAsO, although the magnetic structure is depicted along one axis, that was a representational solution, and both the moment direction and the antiferromagnetic wavevector direction in the FeAs plane along the $a$ or $b$ axis were not determined due to weaker magnetic intensity\cite{A040795}. For BaFe$_2$As$_2$, our powder neutron data unequivocally determine the ferromagnetic row of Fe magnetic moments as along the shorter $b$ axis, and these rows of Fe moments aligning antiferromagnetically along the longer $a$ axis in the FeAs plane (see Fig.~3 and Supplementary Material). This is consistent with the Pauli principle for overlapping $d$-orbitals of neighboring Fe ions, and confirms previous first-principle calculations\cite{A042252}. The Fe moments direct along the longer $a$ axis.

BaFe$_2$As$_2$ differs most significantly from LaFeAsO in that the antiferromagnetic transition occurs at a temperature that is indistinguishable from the structural transition, in contrast to all previously measured FeAs based 1111-type materials where structural and magnetic transitions are separate ones\cite{A040795,A043569,A061450,A062195}. Fig.~4 shows the temperature dependence of the magnetic Bragg peak (101)$_M$, measured at the higher flux triple-axis spectrometer BT7 at NCNR. The solid line represents the mean-field theoretical fit for the squared magnetic order parameter and the N\'{e}el temperature is determined at $T_N=143(4)$ K, which is the same value determined for the structural transition. It is clear from these measurements, that the important criteria for the onset of magnetic order is the complete disappearance of the high temperature phase. The additional flexibility afforded in the 122-type structure by the nearest neighbour layers being displaced by (a/2,b/2) from each other clearly allows the structural strain to be relieved over a shorter temperature range than allowed for the single layered 1111-type systems.

Soon after the discovery of the LaFeAs(O,F) superconductor\cite{Kamihara2008}, it was recognized
that there exist nesting Fermi surfaces connected by the wave vector (1/2,1/2,0)$_T$ in the calculated electronic band structures of the parent compound LaFeAsO\cite{A032740,A033236,A033286}, and the resistivity anomaly of LaFeAsO at $\sim$150 K was predicted to be caused by the resulting spin-density-wave (SDW) order\cite{A033426}. Since the electronic states near the Fermi surfaces come predominantly from the five $d$-orbitals of the Fe ions\cite{A030429,A031279,A031282},
the SDW represents an antiferromagnetic order of Fe. The antiferromagnetic spin fluctuations characterized by the SDW wavevector (1/2,1/2,0)$_T$ have been invoked in a large number of theoretical works as the bosonic mode to mediate the Cooper pairs in the FeAs-based new high $T_C$ superconductors. It turns out that the anomaly is instead associated with a structural transition, and the antiferromagnetic transition
was observed at $\sim$137 K in a separated phase-transition\cite{A040795}. Additionally,
the observed magnetic moment of the Fe ion is an order of magnitude weaker than the theoretically predicated $\sim$2.3 $\mu_B$ per Fe\cite{A033236,A033286}.
Frustrating magnetic exchange
interactions have to be invoked to explain the small observed moment in LaFeAsO\cite{A042252,A042480}.  In another 1111-type material, NdFeAsO, for which magnetic structure has been determined, substantially larger moment 0.9$\mu_B$/Fe was observed below 1.96 K in a combined Nd and Fe antiferromagnetic order, and the resistivity anomaly and the structural transition occur at $\sim$140 K\cite{A061450,A062195}. 
While the predication of the resistivity anomaly caused by the SDW transition has not been realized for the intended 1111-type materials, it is interesting that it occurs in BaFe$_2$As$_2$ in the structurally simpler 122-type materials. The concurring first-order structural and magnetic transition
indicates strong coupling between the structural and magnetic order parameters. A similar previous experimental observation has led to the identification of an orbital ordering among degenerate $d$-orbitals as the driving force of a first-order structural transition in the vicinity of a SDW transition\cite{bao96c}, and we note that there are degenerate Fermi sheets from the Fe $d$-orbitals for the FeAs-based materials\cite{A032740,A033236,A033286}. The structural transition thus can play a more active role than merely as a magnetostrictive consequence of the SDW transition. In summary, a consistent picture covering both the 1111 and 122-type materials is that the Fe magnetic order in the common FeAs layer is the same which breaks the tetragonal symmetry. The occurrence of the Fe magnetic order is contingent upon the orthorhombic structural distortion. Like the CuO$_2$ layers in various cuprates, the FeAs layers have been shown to control similar physical processes in both 122 and 1111-type materials.

%
%
%



\bibliography{/home/bao/kept/tex/bib4/FeAs,/home/bao/kept/tex/bib4/frus,/home/bao/kept/tex/bib4/ruth,/home/bao/kept/tex/bib4/mine}

\begin{thebibliography}{10}
\expandafter\ifx\csname url\endcsname\relax
  \def\url#1{\texttt{#1}}\fi
\expandafter\ifx\csname urlprefix\endcsname\relax\def\urlprefix{URL }\fi
\providecommand{\bibinfo}[2]{#2}
\providecommand{\eprint}[2][]{\url{#2}}

\bibitem{A042053}
\bibinfo{author}{Ren, Z.~A.} \emph{et~al.}
\newblock \bibinfo{title}{Superconductivity at 55 K in iron-based F-doped
  layered quaternary compound Sm[O$_{1-x}$F$_x$]FeAs}.
\newblock \emph{\bibinfo{journal}{Chinese Phys. Lett.}}
  \textbf{\bibinfo{volume}{25}}, \bibinfo{pages}{2215} (\bibinfo{year}{2008}).

\bibitem{A042105}
\bibinfo{author}{Liu, R.~H.} \emph{et~al.}
\newblock \bibinfo{title}{Phase Diagram and Quantum Critical Point in Newly
  Discovered Superconductors: SmO$_{1-x}$F$_x$FeAs}.
\newblock \emph{\bibinfo{journal}{arXiv:0804.2105}}  (\bibinfo{year}{2008}).

\bibitem{A042582}
\bibinfo{author}{Ren, Z.-A.} \emph{et~al.}
\newblock \bibinfo{title}{Superconductivity and Phase Diagram in Iron-based
  Arsenic-oxides ReFeAsO$_{1-\delta}$ without Fluorine Doping}.
\newblock \emph{\bibinfo{journal}{arXiv:0804.2582}}  (\bibinfo{year}{2008}).

\bibitem{A044290}
\bibinfo{author}{Wang, C.} \emph{et~al.}
\newblock \bibinfo{title}{Thorium-doping induced superconductivity in
  Gd$_{1-x}$Th$_x$OFeAs}.
\newblock \emph{\bibinfo{journal}{arXiv:0804.4290}}  (\bibinfo{year}{2008}).

\bibitem{Kamihara2008}
\bibinfo{author}{Kamihara, Y.}, \bibinfo{author}{Watanabe, T.},
  \bibinfo{author}{Hirano, M.} \& \bibinfo{author}{Hosono, H.}
\newblock \bibinfo{title}{Iron-Based Layered Superconductor
  La[O$_{1-x}$F$_x$]FeAs ($x$=0.05 - 0.12) with $T_c$=26K}.
\newblock \emph{\bibinfo{journal}{J.\ Am.\ Chem.\ Soc.}}
  \textbf{\bibinfo{volume}{130}}, \bibinfo{pages}{3296} (\bibinfo{year}{2008}).

\bibitem{A033603}
\bibinfo{author}{Chen, X.~H.} \emph{et~al.}
\newblock \bibinfo{title}{Superconductivity at 43 K in SmFeAsO$_{1-x}$F$_x$}.
\newblock \emph{\bibinfo{journal}{Nature}} \textbf{\bibinfo{volume}{453}},
  \bibinfo{pages}{761} (\bibinfo{year}{2008}).

\bibitem{A033790}
\bibinfo{author}{Chen, G.~F.} \emph{et~al.}
\newblock \bibinfo{title}{Superconductivity at 41 K and its competition with
  spin-density-wave instability in layered CeO$_{1-x}$F$_x$FeAs}.
\newblock \emph{\bibinfo{journal}{Phys. Rev. Lett.}}
  \textbf{\bibinfo{volume}{100}}, \bibinfo{pages}{247002}
  (\bibinfo{year}{2008}).

\bibitem{A034234}
\bibinfo{author}{Ren, Z.~A.} \emph{et~al.}
\newblock \bibinfo{title}{Superconductivity in iron-based F-doped layered
  quaternary compound Nd[O$_{1-x}$F$_x$]FeAs}.
\newblock \emph{\bibinfo{journal}{Europhys. Lett.}}
  \textbf{\bibinfo{volume}{82}}, \bibinfo{pages}{57002} (\bibinfo{year}{2008}).

\bibitem{A034283}
\bibinfo{author}{Ren, Z.~A.} \emph{et~al.}
\newblock \bibinfo{title}{Superconductivity at 52 K in iron-based F-doped
  layered quaternary compound Pr[O$_{1-x}$F$_x$]FeAs}.
\newblock \emph{\bibinfo{journal}{arXiv:0803.4283}}  (\bibinfo{year}{2008}).

\bibitem{A034384}
\bibinfo{author}{Chen, G.~F.} \emph{et~al.}
\newblock \bibinfo{title}{Element substitution effect in transition metal
  oxypnictide Re(O$_{1-x}$F$_x$)TAs (Re=rare earth, T=transition metal)}.
\newblock \emph{\bibinfo{journal}{Chinese Phys. Lett.}}
  \textbf{\bibinfo{volume}{25}}, \bibinfo{pages}{2235} (\bibinfo{year}{2008}).

\bibitem{A043727}
\bibinfo{author}{Ren, Z.~A.} \emph{et~al.}
\newblock \bibinfo{title}{Superconductivity at 53.5 K in GdFeAsO$_{1-\delta}$}.
\newblock \emph{\bibinfo{journal}{Supercond. Sci. Technol.}}
  \textbf{\bibinfo{volume}{21}}, \bibinfo{pages}{082001}
  (\bibinfo{year}{2008}).

\bibitem{A060926}
\bibinfo{author}{Bos, J.-W.~G.} \emph{et~al.}
\newblock \bibinfo{title}{High pressure synthesis of late rare earth
  $R$FeAs(O,F) superconductors; $R$ = Tb and Dy}.
\newblock \emph{\bibinfo{journal}{arXiv:0806.0926}}  (\bibinfo{year}{2008}).

\bibitem{A030429}
\bibinfo{author}{Singh, D.~J.} \& \bibinfo{author}{Du, M.-H.}
\newblock \bibinfo{title}{LaFeAsO$_{1-x}$F$_x$: A low carrier density
  superconductor near itinerant magnetism}.
\newblock \emph{\bibinfo{journal}{Phys. Rev. Lett.}}
  \textbf{\bibinfo{volume}{100}}, \bibinfo{pages}{237003}
  (\bibinfo{year}{2008}).

\bibitem{A031279}
\bibinfo{author}{Haule, K.}, \bibinfo{author}{Shim, J.~H.} \&
  \bibinfo{author}{Kotliar, G.}
\newblock \bibinfo{title}{Correlated electronic structure of
  LaFeAsO$_{1-x}$F$_x$}.
\newblock \emph{\bibinfo{journal}{Phys. Rev. Lett.}}
  \textbf{\bibinfo{volume}{100}}, \bibinfo{pages}{226402}
  (\bibinfo{year}{2008}).

\bibitem{A031282}
\bibinfo{author}{Xu, G.} \emph{et~al.}
\newblock \bibinfo{title}{Doping-dependent Phase Diagram of LaOMAs (M=V-Cu) and
  Electron-type Superconductivity near Ferromagnetic Instability}.
\newblock \emph{\bibinfo{journal}{arXiv:0803.1282 (Europhys. Lett. in-press)}}
  (\bibinfo{year}{2008}).

\bibitem{A054630}
\bibinfo{author}{Rotter, M.}, \bibinfo{author}{Tegel, M.} \&
  \bibinfo{author}{Johrendt, D.}
\newblock \bibinfo{title}{Superconductivity at 38 K in the iron arsenide
  (Ba$_{1-x}$K$_x$)Fe$_2$As$_2$}.
\newblock \emph{\bibinfo{journal}{arXiv:0805.4630}}  (\bibinfo{year}{2008}).

\bibitem{A061209}
\bibinfo{author}{Chen, G.~F.} \emph{et~al.}
\newblock \bibinfo{title}{Superconductivity in hole-doped
  (Sr$_{1-x}$K$_x$)Fe$_2$As$_2$}.
\newblock \emph{\bibinfo{journal}{arXiv:0806.1209}}  (\bibinfo{year}{2008}).

\bibitem{A061301}
\bibinfo{author}{Sasmal, K.} \emph{et~al.}
\newblock \bibinfo{title}{Superconductivity up to 37 K in
  (A$_{1-x}$Sr$_x$)Fe2As2 with A = K and Cs}.
\newblock \emph{\bibinfo{journal}{arXiv:0806.1301}}  (\bibinfo{year}{2008}).

\bibitem{A061459}
\bibinfo{author}{Wu, G.} \emph{et~al.}
\newblock \bibinfo{title}{Transport properties and superconductivity in
  Ba$_{1-x}$M$_x$Fe$_2$As$_2$ (M=La and K) with double FeAs layers}.
\newblock \emph{\bibinfo{journal}{arXiv:0806.1459}}  (\bibinfo{year}{2008}).

\bibitem{A040795}
\bibinfo{author}{Cruz, C.} \emph{et~al.}
\newblock \bibinfo{title}{Magnetic order close to superconductivity in the
  iron-based layered La(O$_{1-x}$F$_x$)FeAs systems}.
\newblock \emph{\bibinfo{journal}{Nature}} \textbf{\bibinfo{volume}{453}},
  \bibinfo{pages}{899} (\bibinfo{year}{2008}).

\bibitem{A043569}
\bibinfo{author}{Nomura, T.} \emph{et~al.}
\newblock \bibinfo{title}{Crystallographic Phase Transition and High-Tc
  Superconductivity in LaOFeAs:F}.
\newblock \emph{\bibinfo{journal}{arXiv:0804.3569}}  (\bibinfo{year}{2008}).

\bibitem{A061450}
\bibinfo{author}{Bos, J.-W.~G.} \emph{et~al.}
\newblock \bibinfo{title}{Influence of the Nd$^{3+}$ Moments on the Magnetic
  Behaviour of the Oxypnictides superconductors NdFeAsO$_{1-x}$F$_x$}.
\newblock \emph{\bibinfo{journal}{arXiv:0806.1450}}  (\bibinfo{year}{2008}).

\bibitem{A062195}
\bibinfo{author}{Qiu, Y.} \emph{et~al.}
\newblock \bibinfo{title}{The absence of the spin-density-wave order in the
  NdFeAsO$_{1-x}$F$_x$ high Tc superconductor system}.
\newblock \emph{\bibinfo{journal}{arXiv:0806.2195}}  (\bibinfo{year}{2008}).

\bibitem{A054021}
\bibinfo{author}{Rotter, M.} \emph{et~al.}
\newblock \bibinfo{title}{Spin density wave anomaly at 140 K in the ternary
  iron arsenide BaFe$_2$As$_2$}.
\newblock \emph{\bibinfo{journal}{arXiv:0805.4021}}  (\bibinfo{year}{2008}).

\bibitem{A061043}
\bibinfo{author}{Krellner, C.} \emph{et~al.}
\newblock \bibinfo{title}{Magnetic and structural transitions in layered FeAs
  systems: AFe$_2$As$_2$ versus RFeAsO compounds}.
\newblock \emph{\bibinfo{journal}{arXiv:0806.1043}}  (\bibinfo{year}{2008}).

\bibitem{gsas}
\bibinfo{author}{{A. Larson and R.B. Von Dreele, {\it GSAS: Generalized
  Structure Analysis System}, (1994)}}.

\bibitem{A042252}
\bibinfo{author}{Yildirim, T.}
\newblock \bibinfo{title}{Origin of the 150 K Anomaly in LaOFeAs; Competing
  Antiferromagnetic Superexchange Interactions, Frustration, and Structural
  Phase Transition}.
\newblock \emph{\bibinfo{journal}{arXiv:0804.2252 (Phys. Rev. Lett. in press)}}
   (\bibinfo{year}{2008}).

\bibitem{A032740}
\bibinfo{author}{Mazin, I.~I.}, \bibinfo{author}{Singh, D.~J.},
  \bibinfo{author}{Johannes, M.~D.} \& \bibinfo{author}{Du, M.~H.}
\newblock \bibinfo{title}{Unconventional sign-reversing superconductivity in
  LaFeAsO$_{1-x}$F$_x$}.
\newblock \emph{\bibinfo{journal}{arXiv:0803.2740}}  (\bibinfo{year}{2008}).

\bibitem{A033236}
\bibinfo{author}{Cao, C.}, \bibinfo{author}{Hirschfeld, P.~J.} \&
  \bibinfo{author}{Cheng, H.~P.}
\newblock \bibinfo{title}{Coexistance of antiferromagnetism with
  superconductivity in LaFeAsO$_{1-x}$F$_x$: effective Hamiltonian from ab
  initio studies}.
\newblock \emph{\bibinfo{journal}{arXiv:0803.3236}}  (\bibinfo{year}{2008}).

\bibitem{A033286}
\bibinfo{author}{Ma, F.} \& \bibinfo{author}{Lu, Z.~Y.}
\newblock \bibinfo{title}{Iron-based layered superconductor
  LaFeAsO$_{1-x}$F$_x$: an antiferromagnetic semimetal}.
\newblock \emph{\bibinfo{journal}{arXiv:0803.3286}}  (\bibinfo{year}{2008}).

\bibitem{A033426}
\bibinfo{author}{Dong, J.} \emph{et~al.}
\newblock \bibinfo{title}{Competing Orders and Spin-Density-Wave Instability in
  La(O$_{1-x}$F$_x$)FeAs}.
\newblock \emph{\bibinfo{journal}{arXiv:0803.3426 (Europhys. Lett. in-press)}}
  (\bibinfo{year}{2008}).

\bibitem{A042480}
\bibinfo{author}{Si, Q.} \& \bibinfo{author}{Abrahams, E.}
\newblock \bibinfo{title}{Strong Correlations and Magnetic Frustration in the
  High Tc Iron Pnictides}.
\newblock \emph{\bibinfo{journal}{arXiv:0804.2480}}  (\bibinfo{year}{2008}).

\bibitem{bao96c}
\bibinfo{author}{Bao, W.} \emph{et~al.}
\newblock \bibinfo{title}{Dramatic switching of magnetic exchange in a classic
  transition metal oxide: evidence for orbital ordering}.
\newblock \emph{\bibinfo{journal}{Phys. Rev. Lett.}}
  \textbf{\bibinfo{volume}{78}}, \bibinfo{pages}{507} (\bibinfo{year}{1997}).

\end{thebibliography}


\begin{addendum}
 \item Work at LANL is supported by U.S.\ DOE, at USTC by the Natural Science Foundation of China, Ministry of Science and Technology of China (973 Project No: 2006CB601001) and by National Basic Research Program of China (2006CB922005).
 \item[Author information] The authors declare that they have no
competing financial interests.
 Correspondence and requests for materials
should be addressed to W.B. (wbao@lanl.gov).
\end{addendum}


\newpage

\includegraphics[width=18cm,angle=0,clip=true]{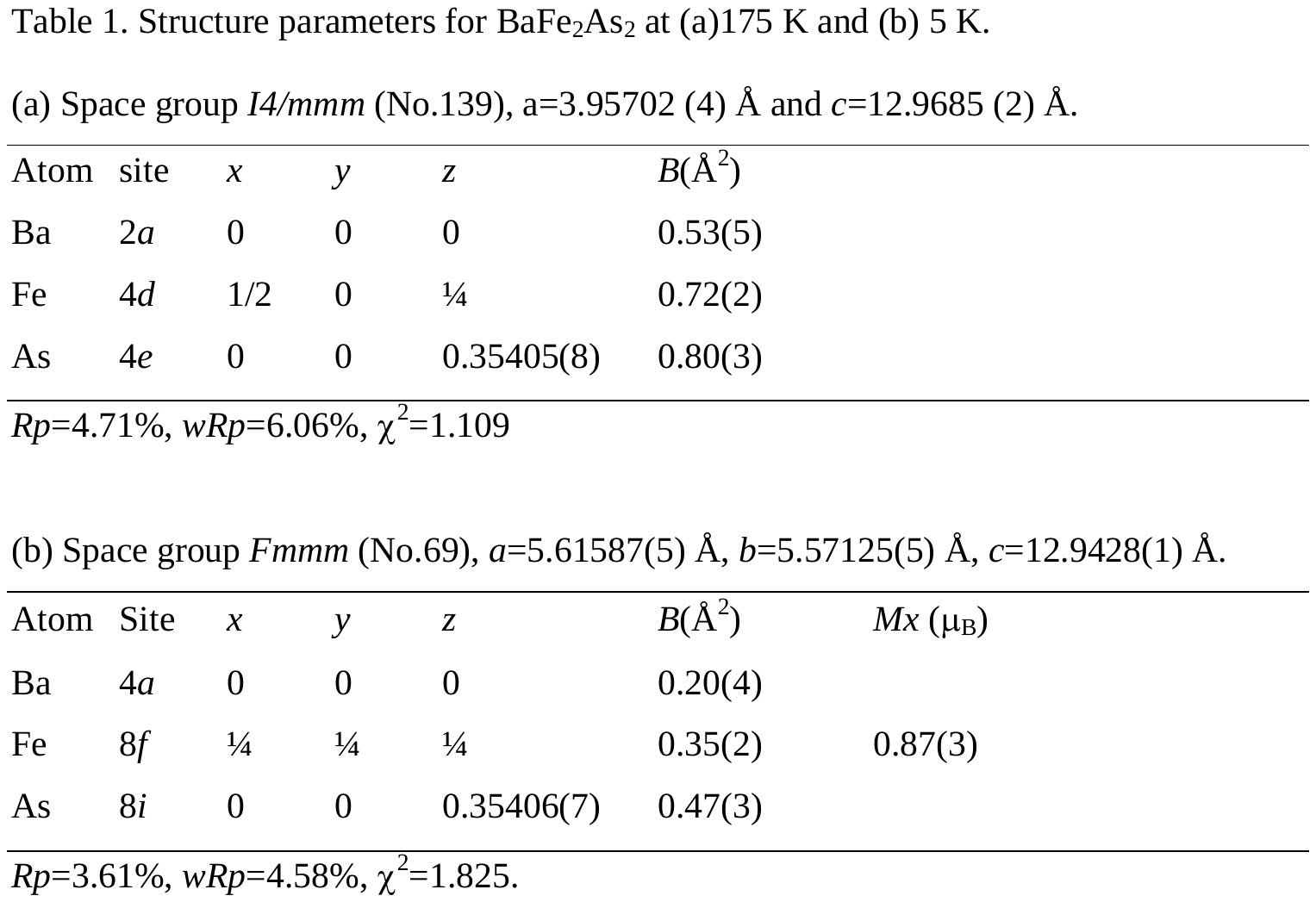}

\newpage
\begin{figure}
\includegraphics[width=16cm,angle=0,clip=true]{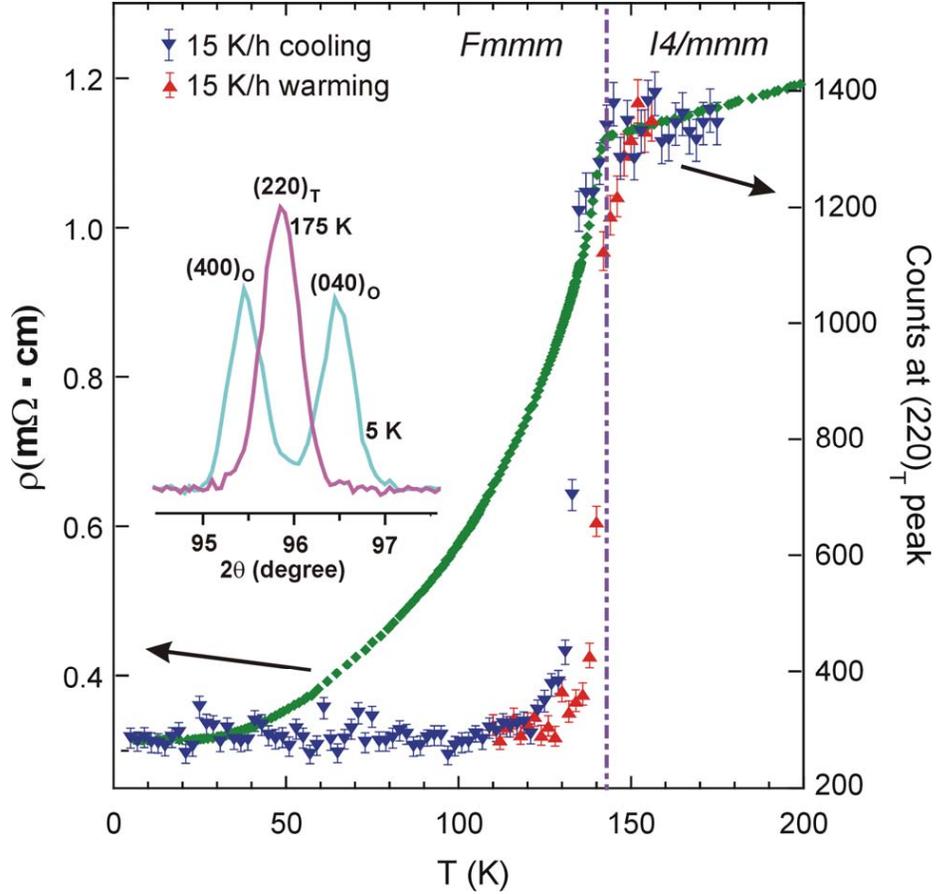}
\vskip -5.5cm
\caption{The resistivity of BaFe$_2$As$_2$ as a function of temperature, measured while cooling, shows a sharp drop at $T_S\approx 142$ K. The resistivity anomaly is associated with a first-order structural transition, as indicated by the neutron diffraction intensity at diffraction angle $2\theta=95.8^o$. Inset: Neutron diffraction spectra measured above and below the structural transition, showing the splitting of the peak below $T_S$.}
\label{fig1}
\end{figure}

\newpage

\begin{figure}
\includegraphics[width=15cm,angle=0,clip=true]{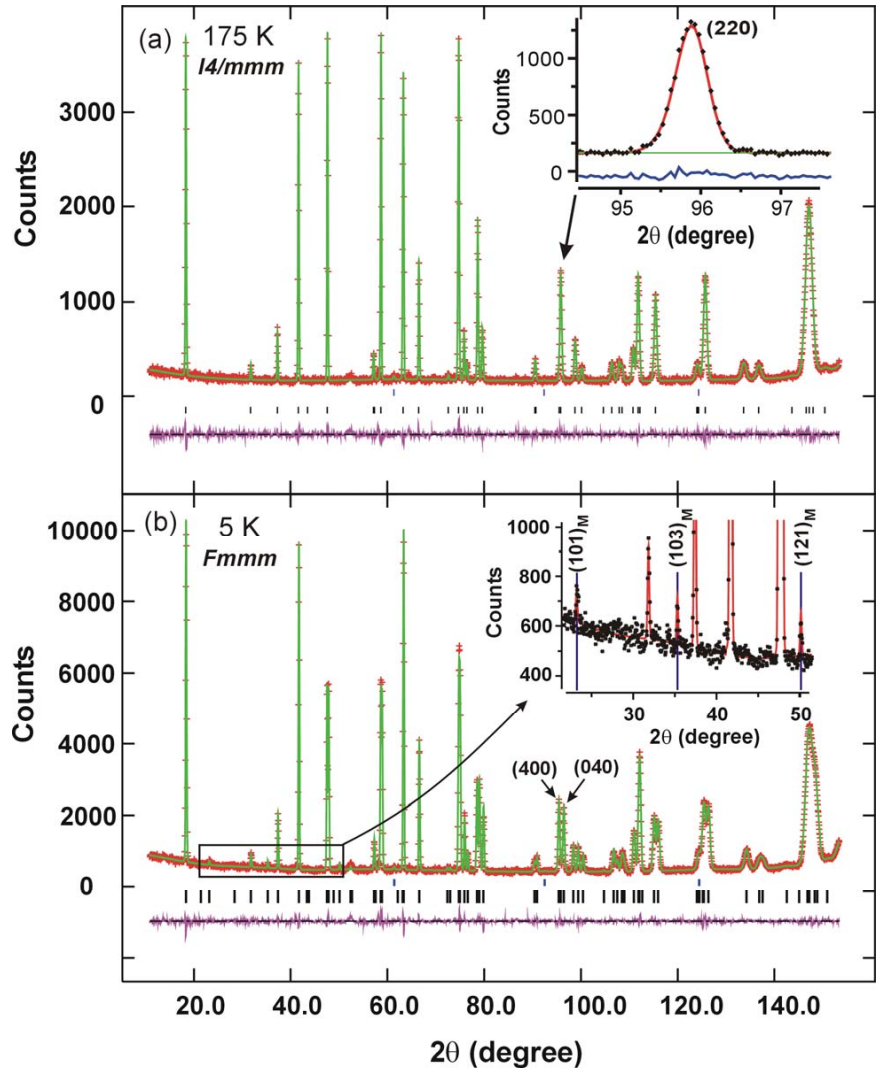}
\vskip -3.7cm
\caption{Neutron powder diffraction spectra at (a) 175 and (b) 5 K. The high temperature spectrum is refined with the tetragonal $I4/mmm$ space group; the low temperature one with the orthorhombic $Fmmm$ space group together with the magnetic structure shown in Fig. 3. The (220) Bragg peak in (a) is split into the (400) and (040) in (b) by the orthorhombic structural distortion. Magnetic peaks at 5 K are highlighted in the inset to (b).}
\end{figure}

\newpage

\begin{figure}
\includegraphics[width=15cm,angle=0,clip=true]{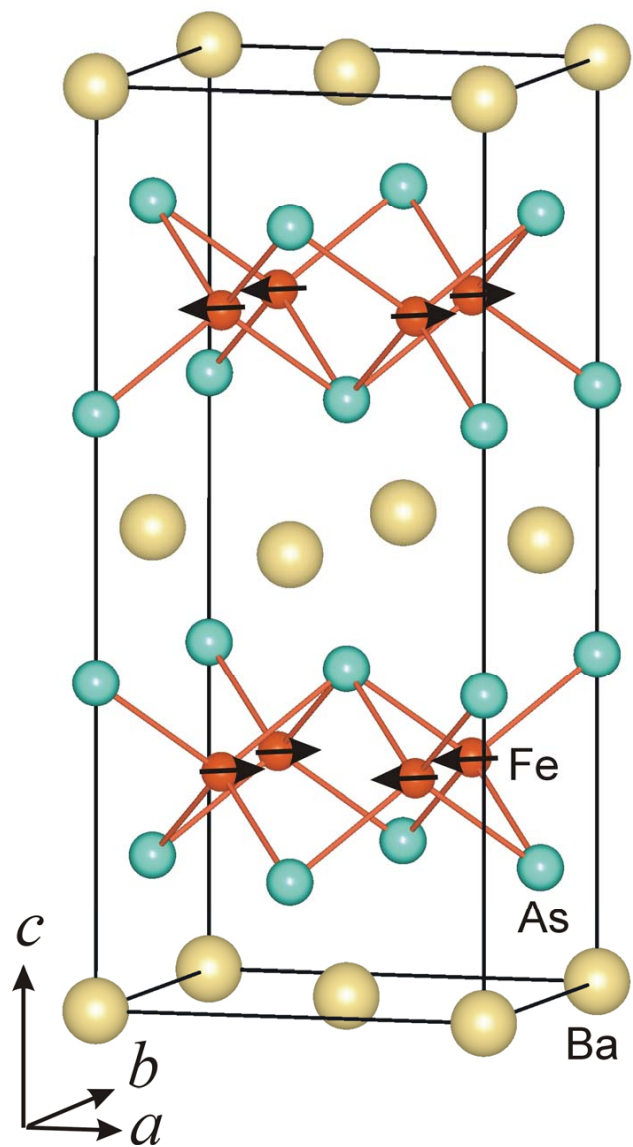}
\vskip -3cm
\caption{Magnetic and crystal structures of BaFe$_2$As$_2$ shown in an orthorhombic $Fmmm$ unit cell.}
\end{figure}

\newpage

\begin{figure}
\includegraphics[width=15cm,angle=0,clip=true]{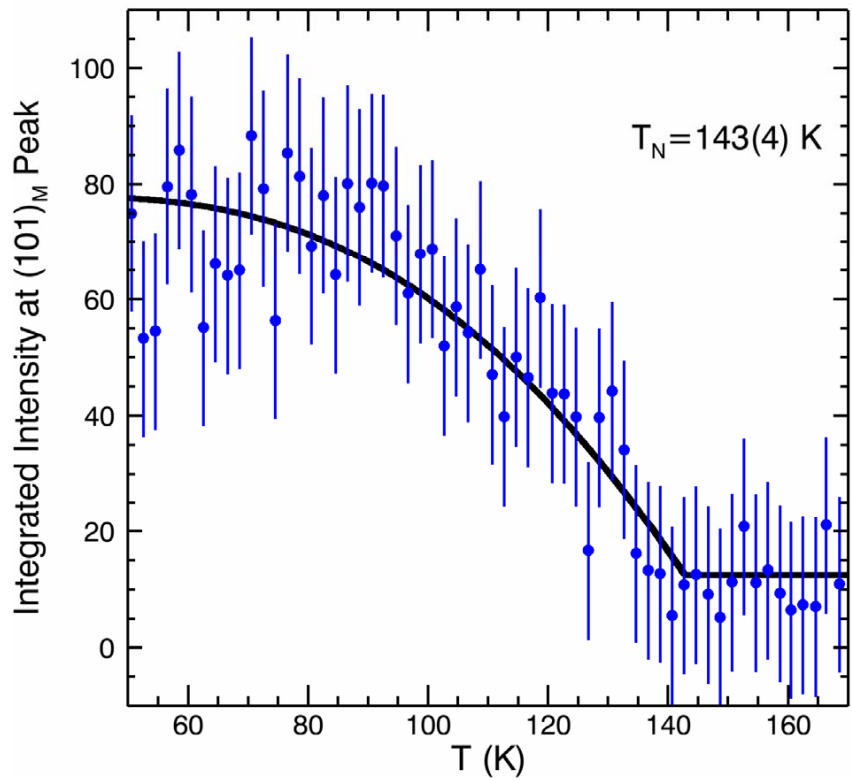}
\vskip -3cm
\caption{Magnetic Bragg peak (101)$_M$ as a function of temperature. The solid line represents the least-square fit to mean-field theory for the squared order-parameter.}
\end{figure}

\section*{Supplementary Information}

\setcounter{equation}{0}
\renewcommand{\theequation}{S\arabic{equation}}

\noindent{\bf Determination of the magnetic wavevector and magnetic moment direction in BaFe$_2$As$_2$}

The nesting wavevector (1/2,1/2,0)$_T$ of the calculated Fermi surface from the FeAs square-lattice layer\cite{A032740,A033236,A033286} has a symmetry related twin (1/2,-1/2,0)$_T$. Therefore, in principle, either (100)$_O$, (010)$_O$, or both could be the in-plane component of the magnetic wavevector characterizing the antiferromagnetic structure in the orthorhombic phase of BaFe$_2$As$_2$. Figure S1 shows that the difference between the lattice parameter $a$ and $b$ is large enough for the measured magnetic Bragg peak to be indexed as the (121) instead of (211). Therefore, the magnetic wavevector is (101), namely, the Fe magnetic moments are aligned antiferromagnetically along the $a$ and $c$-axis, and ferromagnetically along the $b$-axis.

The in-plane magnetic moment has a two-fold symmetry [Wang, X. F. et al., arXiv:0806.2452 (2008)], namely, it is either along the $a$ or $b$-axis of the $Fmmm$ unit cell. Refinement yields that the magnetic moment is parallel to the $a$-axis, as presented in 
the bottom row of Fig.~S1 and in Table 1(b).

\newpage

\setcounter{figure}{0}
\renewcommand{\thefigure}{Figure S\arabic{figure}}
\begin{figure}
\includegraphics[width=15cm,angle=0,clip=true]{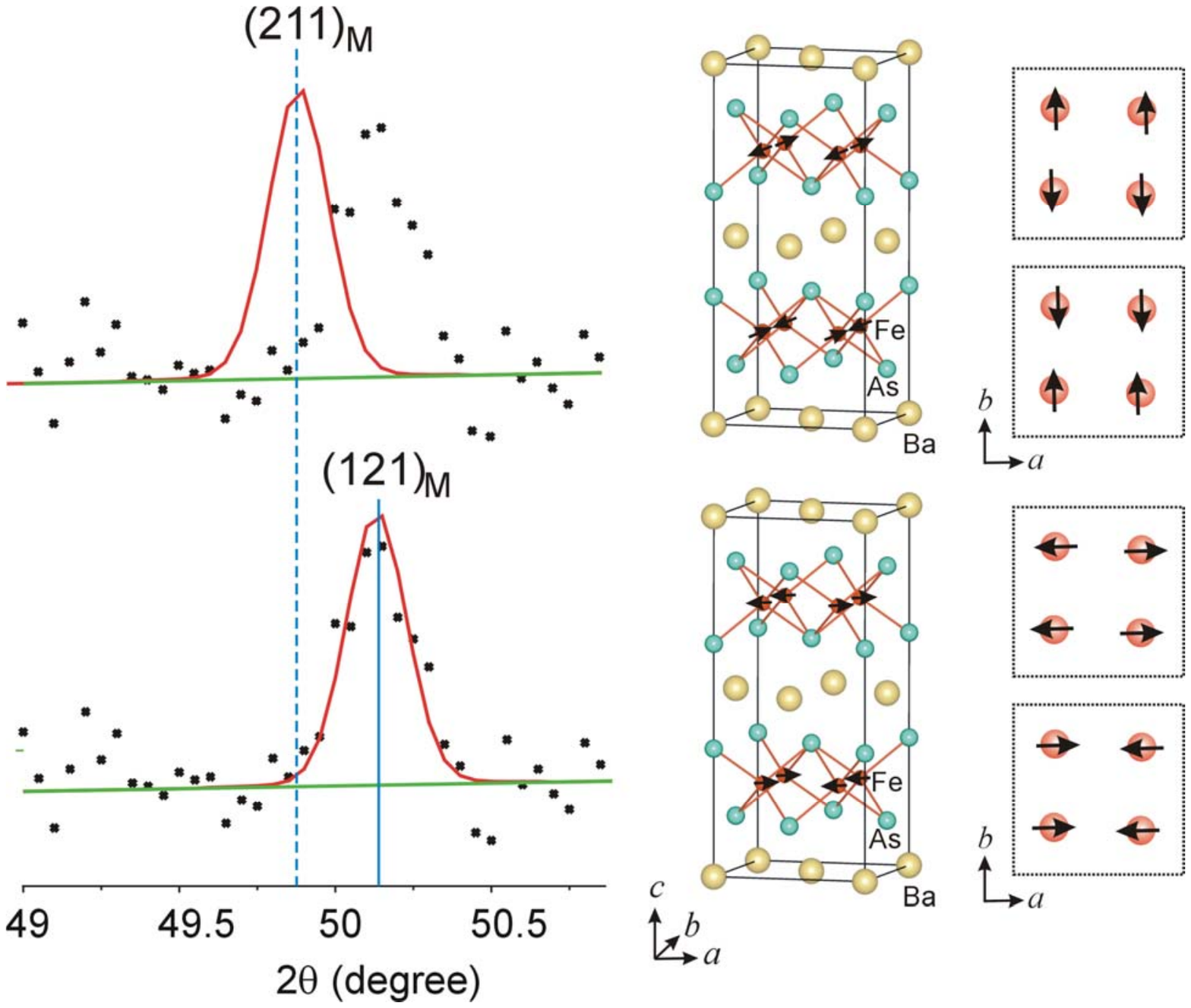}
\vskip -3cm
\caption{{\bf Comparison of two magnetic models}. The measured magnetic Bragg peak is well represented by the (121) index. The (211) peak position is outside the measured peak. The corresponding magnetic model for each row is shown on the right in the orthorhombic unit cell and in the two Fe layers within the unit cell.}
\end{figure}

\end{document}